\documentclass{article}

\usepackage{PRIMEarxiv}

\usepackage[utf8]{inputenc} % allow utf-8 input
\usepackage[T1]{fontenc}    % use 8-bit T1 fonts
\usepackage{hyperref}       % hyperlinks
\usepackage{url}            % simple URL typesetting
\usepackage{booktabs}       % professional-quality tables
\usepackage{amsfonts}       % blackboard math symbols
\usepackage{nicefrac}       % compact symbols for 1/2, etc.
\usepackage{microtype}      % microtypography
\usepackage{graphicx}       % graphics
\usepackage{svg}
\usepackage{float}
\usepackage{subcaption}
\usepackage{textgreek}
\usepackage{amsmath} 

\graphicspath{{Figures/}}

\title{Accurate and Noise-Robust Wavefront Reconstruction with an Optical Vortex Wavefront Sensor}

\author{
  \textbf{Magdalena Łukowicz}$^{1}$, \textbf{Aleksandra K. Korzeniewska}$^{1}$, \textbf{Kamil Kalinowski}$^{1}$,\\
  \textbf{Rafał Cichowski}$^{1}$, \textbf{Rosario Porras-Aguilar}$^{2}$, and \textbf{Mateusz Szatkowski}$^{1,*}$\\
  \\
  $^{1}$Wroc\l{}aw University of Science and Technology, Department of Optics and Photonics,\\
  Wybrzeże Wyspiańskiego 27, 50-370 Wrocław, Poland\\
  $^{2}$The University of North Carolina at Charlotte, Department of Physics and Optical Science,\\
  Charlotte, North Carolina, 28223, USA\\
  $^{*}$\texttt{mateusz.szatkowski@pwr.edu.pl}
}

\begin{document}
\maketitle

\begin{abstract}
The term wavefront sensor refers to the entire class of devices capable of measuring the optical wavefront of the incoming beam. Although numerous solutions have been proposed so far, recent advances in structured light have opened new development possibilities through controlled modification of optical field amplitude and phase. We present an alternative approach to angle-based sensing, introducing optical vortices—stable phase singularities—within each subaperture of the Shack-Hartmann (S-H) architecture. Rather than changing the fundamental angle-based operating principle, it transforms the tracked quantity and its detection method. The presence of a singularity enables a dedicated tracking algorithm that outperforms conventional methods without increasing computational complexity. We evaluated its performance against the conventional S-H across a broad SNR range (from 2 to 22), corresponding to shot-noise limited to high saturation regimes, demonstrating lower mean residual phase variance across all conditions. This work demonstrates that structured beam shaping can extend the capabilities of traditional S-H architectures without requiring fundamental redesign.
\end{abstract}

\keywords{Optical vortices \and phase singularities \and wavefront sensing \and Shack-Hartmann sensor \and adaptive optics}

\section{Introduction}

Wavefront sensors are traditionally divided into two main classes. The first class comprises interferometry and diffractive optics. These methods estimate the wavefront using intensity modulation given by interference patterns, diffraction effects, or intensity variations between different optical planes \cite{Medecki1996, Ghebremichael2008, Wu2019, Wu2021b, Khorin2025}. The other class covers angle-based sensors, which measure the spatial distribution of the incident wavefront, locally sectioning its structure to estimate local angles \cite{Platt2001, Esposito2001, Ragazzoni2002}. 

Both sensor classes have demonstrated their capabilities across a vast spectrum of fields. They compensate for atmospheric turbulence in astronomy \cite{Bolbasova:19}, measure optical aberrations in artificial optical systems \cite{rahman-2013}, enable novel imaging in vision optics \cite{Brunner2021}, perform precision metrology for freeform optics manufacturing \cite{Swain2019}, and offer modal decomposition \cite{Anguita2022}. This range of applications is expected to expand due to the incorporation of deep learning approaches \cite{Nishizaki2019}.

Many of these fields and their proposed wavefront sensing solutions relied on a few primary sensor architectures. One fundamental example is the Shack-Hartmann wavefront sensor \cite{Platt2001, WU2023107289}. Its working principle follows the angle-based methodology. It consists of a lenslet array and a CCD detector. The sensor reconstructs the incoming wavefront by measuring the displacement of each focal spot from its on-axis position, providing the local wavefront slope. All calculated slopes are later integrated over the entire pupil, enabling wavefront retrieval.

However, modern laser beam shaping opportunities have opened new development possibilities. Spatial light modulators (SLMs) can now modify the complex amplitude of optical fields, enabling modifications to traditional sensor architectures and their working principles. Recent innovations have focused on enhancing the conventional S-H architecture by modifying the phase structure within each sub-aperture through SLMs.

An interesting early example of such an active sensor was proposed by Bowman et al., who transformed the conventional Shack-Hartmann sensor into its holographic version using an SLM \cite{Bowman2010}. This supported the authors' research on optical trapping, as it enabled the identification and correction of aberrations within the optical trapping setup. Building on this concept, in \cite{Lechner2020}, the authors proposed an adaptable Shack-Hartmann sensor that adjusted the size of microlenses by redesigning the phase map displayed on the SLM, to mitigate the scintillation effect, showing that an increase in lens diameter can efficiently reduce these unwanted intensity fluctuations. Taking a different approach, \cite{Dubey2021} developed a COACH-based Shack–Hartmann wavefront sensor with an array of phase-coded masks. The lenslet array of S-H was replaced with an array of coded phase masks to lower the mean square wavefront error by an order of magnitude in comparison to the conventional S-H. Extending these capabilities to multi-object sensing, \cite{Singh2025} proposed a novel concept to separately measure the effect of atmospheric turbulence on three distinct stars. The diffractive structure displayed on the SLM separated the incoming light, thus each star could be measured separately using the Shack-Hartmann angle-based principles. Most recently, \cite{Wu2025} introduced novel beam geometries into each subaperture and improved the dynamic range of the S-H wavefront sensor. This work showed that one could take advantage of complex beam shaping within each subaperture.

Based on these developments and inspired by the potential for complex beam shaping within subapertures, we propose a new holographic wavefront sensor: the Optical Vortex Wavefront Sensor (OVWS). We modified the conventional Shack-Hartmann (S-H) design by introducing an optical vortex into each subaperture. This combines the adaptability provided by SLMs with the metrological features of optical vortices - stable phase singularities in the optical field, manifesting as dark points in the intensity distribution of the optical field \cite{gbur-2016}. 

The proposed sensor falls between interferometry and angle-based wavefront retrieval principles, utilizing the angle-based principle but relying on the interferometric phenomenon of phase singularity. We estimate the incoming wavefront by measuring the displacement of the singular point. The existence of a phase singularity reverses the concept of spot tracking, where instead of a bright spot, we track the zero-intensity point using the dedicated localisation algorithm relying on the \textit{a priori} information about the existence of singular points. We enhanced the localization precision in the presence of shot noise compared to the conventional S-H sensor, providing a wavefront sensing solution in the low-light regime, being a challenging domain for traditional sensors. In contrast to neural network approaches \cite{Gu_2021, Li:18} and computational filtering methods \cite{Gu2022} addressing the issue of wavefront retrieval in the low-light regime, our work proposes an all-optical solution that fundamentally changes both the measured quantity (singular point) and the tracking mechanism, without increased computational complexity. This work shows that complex beam shaping could lead to extended capabilities, despite relying on the traditional S-H architecture.

\section{Design}

In the following section, we go over the mathematical framework and implementation approach for the OVWS design. The OVWS implementation relies on introducing controlled phase singularities into each subaperture through careful phase modulation \cite{Rosales-Guzmán_2024}. To assure the beam is modulated correctly, we start by defining the base grating that aims to shift the modulated light (1st diffraction order, containing the sensor's structure) away from its unmodulated part (0th diffraction order). Adjusting the approach presented in \cite{Bowman2010}, the base grating is inserted in the central subaperture. We superpose it with the spiral phase that introduces the optical vortex into the beam. Therefore, the phase inside the base aperture is given by:

\begin{equation}
\label{eq:phibase}
\phi_{\text{base}} = l\theta + 2\pi(f_x x + f_y y)
\end{equation}

\noindent where $l$ determines the optical vortex topological charge, $\theta = \arctan(y/x)$ is the azimuthal angle, and $f_x$, $f_y$ denote grating spatial frequencies in the $x$ and $y$ direction, respectively. 

\begin{equation}
\label{eq:fxy_base}
f_{x,y} = \frac{n_{x,y}}{p_{x,y}}
\end{equation}

\noindent where $n_{x,y}$ denote the number of grooves in both $x$ and $y$ directions, and $p_{x,y}$ denotes the number of SLM's pixels in $x$ and $y$ directions. 

Controlling the gratings' spatial frequency is crucial as it changes the separation between the spots at the image plane. The spatial frequency of each subsequent aperture's grating is dictated by the vector $\mathbf{V}_i=[v_{ix},v_{iy}]$ from the center of the base aperture to the particular aperture $i$, modified by the scaling parameter $s$. The grating frequency magnitude depends on the distance from the center, while the orientation follows the aperture's angular position:

\begin{equation}
\label{eq:fxy_modified}
f_{ix} = \frac{n_x + s \cdot |\mathbf{V}_i| \cdot \cos(\alpha_i)}{p_x}, \quad f_{iy} = \frac{n_y + s \cdot |\mathbf{V}_i| \cdot \sin(\alpha_i)}{p_y}
\end{equation}

\noindent where $|\mathbf{V}_i| = \sqrt{v_{ix}^2 + v_{iy}^2}$ is the distance from center and $\alpha_i = \arctan(v_{iy}/v_{ix})$ is the direction angle.

The scaling parameter $s$ controls spots separation at the Fourier plane. This parameter must be chosen experimentally, taking into account the system's numerical aperture and the detector's size. Combining equations \ref{eq:phibase} and \ref{eq:fxy_modified}, the general equation describing the phase inside each aperture $i$ is obtained:

\begin{equation}
\label{eq:phifinal}
\phi_i = l\theta + 2\pi\left(\frac{n_x + s \cdot |\mathbf{V}_i| \cdot \cos(\alpha_i)}{p_x} x + \frac{n_y + s \cdot |\mathbf{V}_i| \cdot \sin(\alpha_i)}{p_y} y\right)
\end{equation}

where $x$ and $y$ are the pixel coordinates on the SLM. It is worth noticing that equation \ref{eq:phifinal} for $v_{ix} = v_{iy} = 0$ describes the phase inside the base aperture, as defined in \ref{eq:phibase}. In case the $l=0$, the equation \ref{eq:phifinal} describes the conventional S-H, as proposed in \cite{Bowman2010}. Thus, we will continue to design holograms with the method described above, switching from conventional S-H ($l=0$) to OVWS ($l=1$) to compare performance of these two.

\begin{figure}[htbp]
\centering
\includegraphics[width=\textwidth]{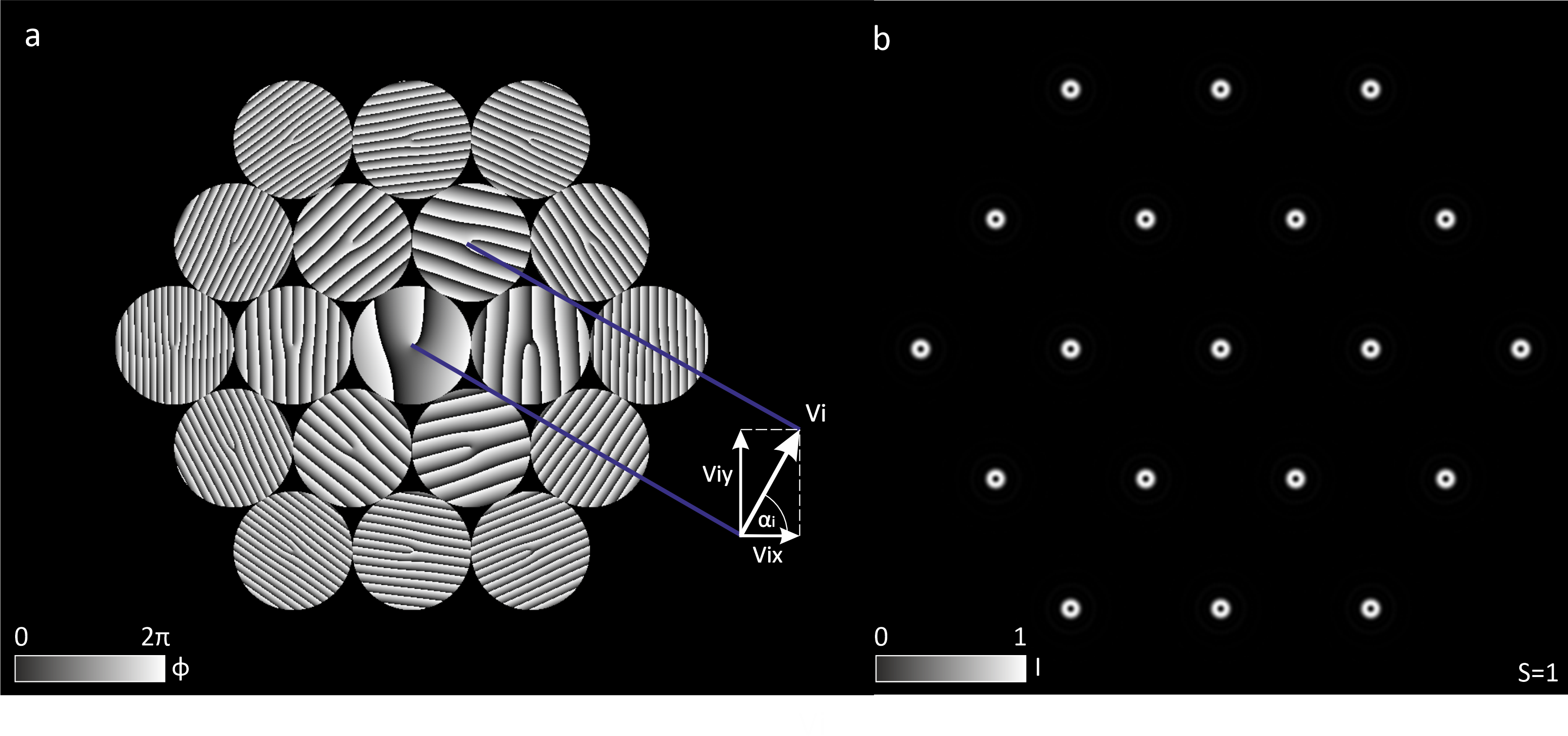}
\caption{a) Concept of the hologram design (spatial frequency of diffraction gratings is not in scale). The base grating is inserted in the central subaperture. Starting from there, vector $\mathbf{V}_i$ sets spatial frequencies $f_{ix}$, $f_{iy}$ in subsequent subapertures following equation \ref{eq:fxy_modified} together with the spiral phase as shown in equation \ref{eq:phifinal}  b) Simulated intensity distribution at the Fourier plane with scaling parameter $s=1$}
\label{fig:Hologram_design}
\end{figure}

The exemplary hologram of the OVWS is presented in Figure \ref{fig:Hologram_design}a. We employed the hexagonal grid, as it provides an equal distance between neighboring spots at the Fourier plane and allows a more uniform sampling than the square one \cite{Gantes-Nuez2018}. It has been shown that in case of S-H architectures, results for the hexagonal and square grids are comparable \cite{Navarro2009}.

Regardless of the chosen grid distribution (hexagonal or square), equation \ref{eq:phifinal} will be maintained, correctly determining the spatial frequencies of the subsequent gratings within each aperture. The simulated intensity distribution at the Fourier plane is shown in Figure \ref{fig:Hologram_design}b. 

Similarly to the conventional S-H wavefront sensor, the optimal number of Zernike modes $Z_{\text{opt}}$ that can be reliably reconstructed depends on the total number of subapertures, represented as $M$ subapertures across the diameter, following \cite{Li2002, Gladysz2022}:

\begin{equation}
Z_{\text{opt}} = \left\lfloor 2.04M^{1.36} \right\rfloor
\end{equation}

where $\lfloor...\rfloor$ denotes the largest integer less than or equal to the value inside.
Although the above equation was derived for a square grid, we apply it consistently here, as in the case of a hexagonal grid with constant N, the number of subapertures is higher than for a square grid. Thus, this limit sets a reasonable boundary on the number of modes $Z_{\text{opt}}$ that the sensor can reconstruct. Therefore, it is essential to accurately define the pupil diameter, which determines the limit for the number of subapertures that fall within it. This has to take into account the dimension and the resolution of the SLM and the diameter of the incoming beam, so that each suberture takes part in the beam shaping process. 

Having established the theoretical framework for OVWS design, we now validate its performance through numerical simulations, focusing on the key advantage of improved noise resistance.

\section{Numerical simulations}

\begin{figure}[htbp]
\centering
\includegraphics[width=280 pt]{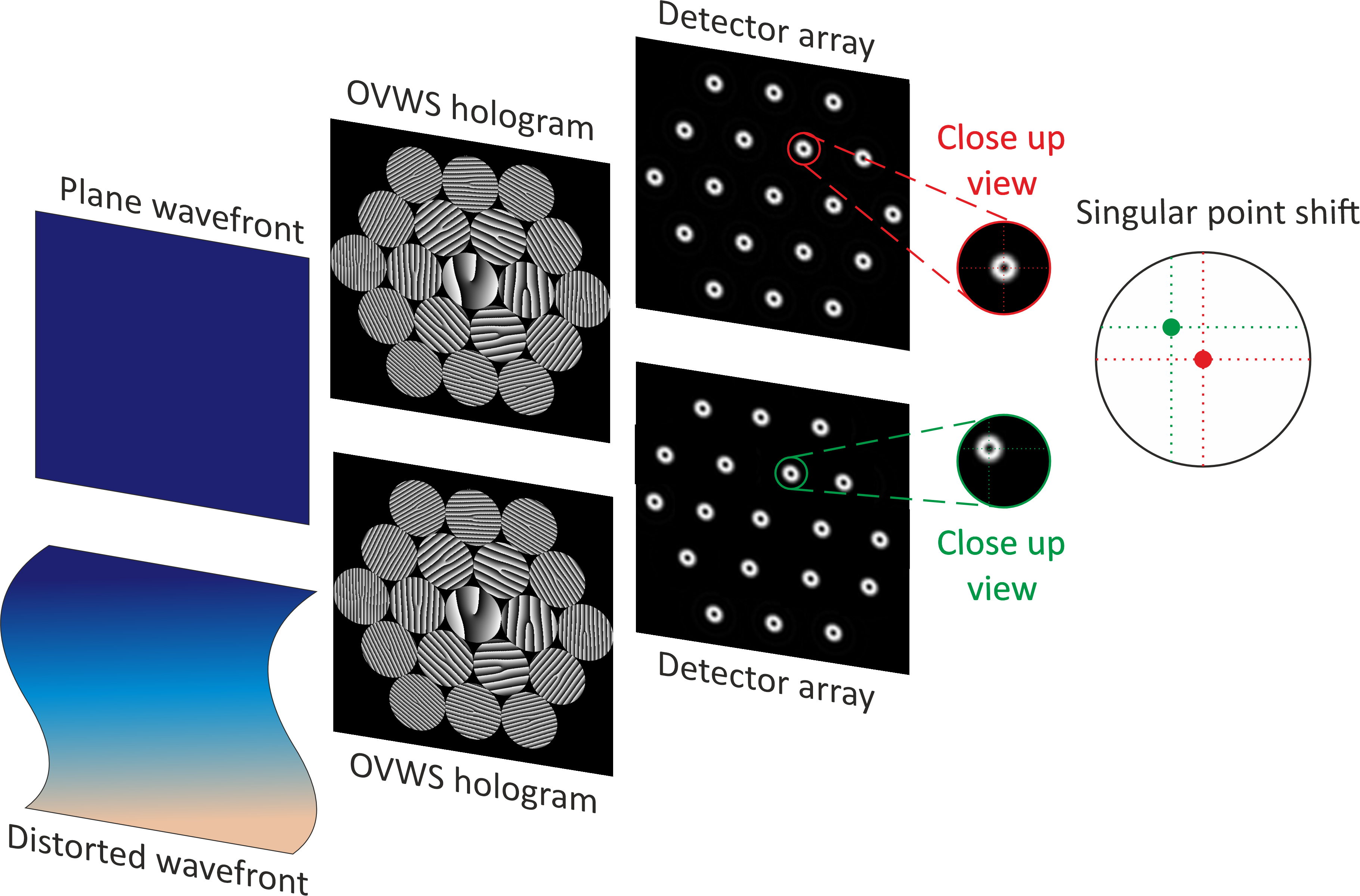}
\caption{Concept of the measurement principle with the OVWS.}
\label{fig:Concept}
\end{figure}

The main concept of the OVWS is analogous to that of the conventional S-H sensor (both evaluate the incoming wavefront through focal spot displacements). However, replacing conventional Gaussian beams with optical vortices fundamentally transforms the retrieval scheme. Instead of localizing bright focal spots (conventional, Gaussian), we determine the position of phase singularities—zero-intensity points within each optical vortex (Figure \ref{fig:Concept}).

To examine localization performance, we compared the precision of both approaches under controlled shot noise. We generated single spots for both configurations: a conventional Gaussian spot for the S-H sensor and an optical vortex spot for the OVWS. Shot noise was then added to each spot type, and localization was performed using method-specific algorithms: weighted centroid \cite{antonello-2014} for the conventional Gaussian spots and the Laguerre-Gaussian transform \cite{Wang2006} for the optical vortex spots, originally developed for speckle metrology but recently adapted for efficient off-axis singularity tracking \cite{Szatkowski2022, Korzeniewska2025}. Both procedures are explained in detail in S1.

Before comparing localization performance, we clarify our terminology. The beam structure of the OVWS at the Fourier plane is a Laguerre-Gaussian mode regardless of the generation method (phase or phase-amplitude modulation) \cite{Vallone2016}. Throughout this work, we use the term "optical vortex" for sensor design and "Laguerre-Gaussian transform" for localization methodology. 

To introduce shot noise, we followed \cite{Konnik2014} and implemented a numerical model that mimics digital detector read noise (the combined output of photon-to-charge, charge-to-voltage, and voltage-to-digital conversion errors). We applied $N=100$ separate random distributions per noise level to single spot images. These spots were simulated numerically by subsequent sensors (conventional S-H and OVWS) with identical parameters. The simulated read noise ranged from 5e to 85e, where the higher level of read noise indicates a more noisy environment, ranging from 22 to 10 signal to noise ratio (SNR), respectively. These range mimics the real experimental scenario \cite{Wei2020}. Further, we localized the position of the spot for each distribution, and calculated the Root Mean Square ($\text{RMS}_{\text{noise}}$) value per noise level:

\begin{equation}
\text{RMS}_{\text{noise}} = \sqrt{\frac{1}{N} \sum_{i=1}^{N} \left[ (x_i - \bar{x})^2 + (y_i - \bar{y})^2 \right]}
\end{equation}

where the 2D centroid $(\bar{x},\bar{y})$ is given by:

\begin{equation}
 \quad \bar{x} = \frac{1}{N} \sum_{i=1}^{N} x_i, \quad \bar{y} = \frac{1}{N} \sum_{i=1}^{N} y_i
\end{equation}

\begin{figure}[htbp]
\centering
\includegraphics[width=\textwidth]{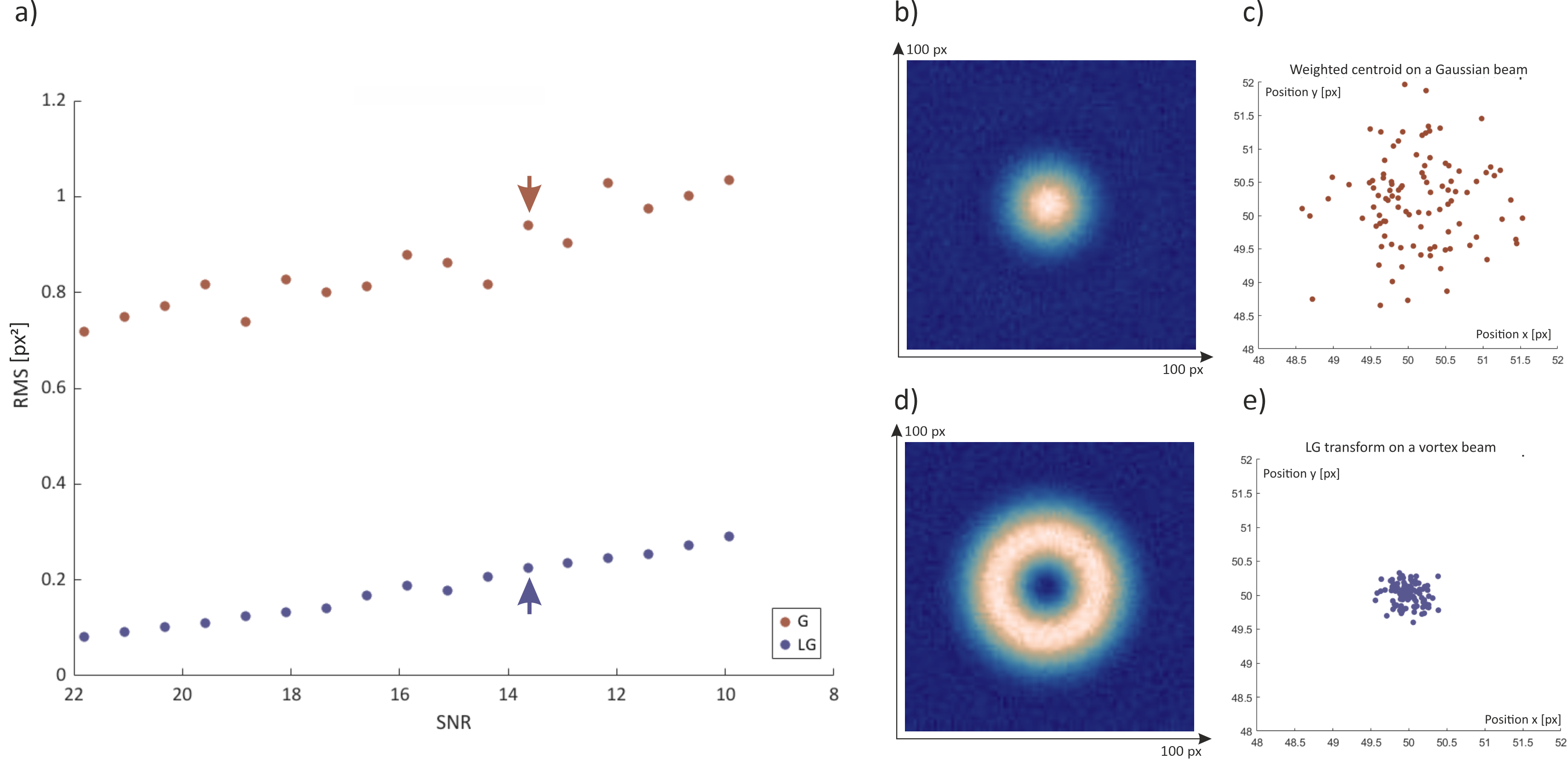}
\caption{a) Localization precision under the manually introduced noise for both the weighted centroid applied to a Gaussian beam (orange) and the Laguerre-Gaussian transform applied to an optical vortex (blue). Arrows indicate the level of the noise shown in subsequent subfigures b) Single iteration of the read noise distribution (13 SNR) for the Gaussian beam c) Scattered localization positions for each iteration of noise presented in b, d) Single iteration of the read noise distribution (13 SNR) for the optical vortex e) Scattered localization positions for each iteration of noise presented in d} 
\label{fig:Shot_noise_single}
\end{figure}

The combined results comparing localization precision are presented in Figure~\ref{fig:Shot_noise_single}a. These averaged results, obtained from 100 independent measurements per noise level, quantitatively demonstrate that, despite the presence of shot noise, the LG transform method applied to vortex spots provides improved noise resistance compared to the conventional centroid localization on the Gaussian beam. The improvement in precision remains consistent across the tested noise range from 22 to 9 SNR, corresponding to 85e to 5e read noise levels), with the LG method maintaining lower RMS reconstruction errors throughout. Exemplary images of noisy spots are shown in Figure~\ref{fig:Shot_noise_single}b and d, while the corresponding point clouds of these 100 localized positions per exemplary noise level are presented in Figure~\ref{fig:Shot_noise_single}c and e. 

These simulations confirm the theoretical advantage of the OVWS localization approach. To validate this performance improvement under realistic wavefront sensing, we conducted numerical wavefront retrieval using 100 field instances $k$ per noise level, retrieving the $\phi_k$ wavefront and calculating the residual phase $\theta_k$ from the numerically introduced aberration $\phi_{\text{introduced}}$:

\begin{equation}
\label{eq:retrieval}
\theta_k = \phi_{\text{introduced}} - \phi_k
\end{equation}

To measure the wavefront retrieval accuracy, the mean residual phase variance $\langle\sigma^2\rangle$  has been calculated, taking into account all residual phases for a given read noise value:

\begin{equation}
\label{eq:residual_variance}
\langle\sigma^2\rangle = \left\langle\sum_{k=1}^{100} \theta_k^2\right\rangle
\end{equation}

\begin{figure}[htbp]
\centering
\includegraphics[width=\textwidth]{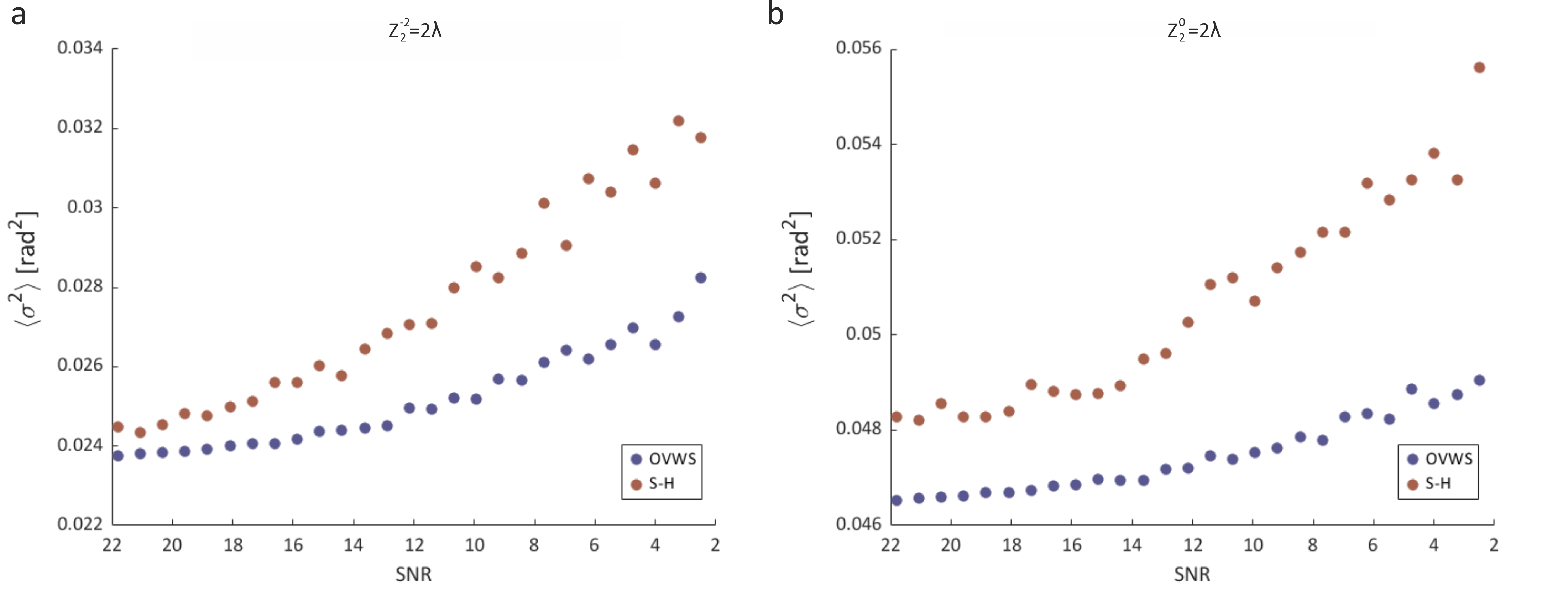}
\caption{Mean residual phase variance $\langle\sigma^2\rangle$ as a function of SNR for: a) Oblique astigmatism $Z^{-2}_2=2\lambda$ b) Defocus $Z^{0}_2=2\lambda$} 
\label{fig:Numerical_wavefront}
\end{figure}

These procedures have been repeated for both S-H and OVWS, and the summarized plots for both astigmatism $Z^{-2}_2=2\lambda$ and defocus $Z^{0}_2=2\lambda$ aberrations (defined over a 1000 px pupil) are shown in Figure \ref{fig:Numerical_wavefront}. For the entire analyzed SNR range, the OVWS provided consistently lower residual phase variance than S-H. This difference increases once the SNR goes to the very low regime, SNR$<$10. Having demonstrated OVWS advantage numerically, we proceeded to experimental validation using controllable noise levels achieved through camera gain modulation.

\section{Experimental results}

\begin{figure}[htbp]
\centering
\includegraphics[width=\textwidth]{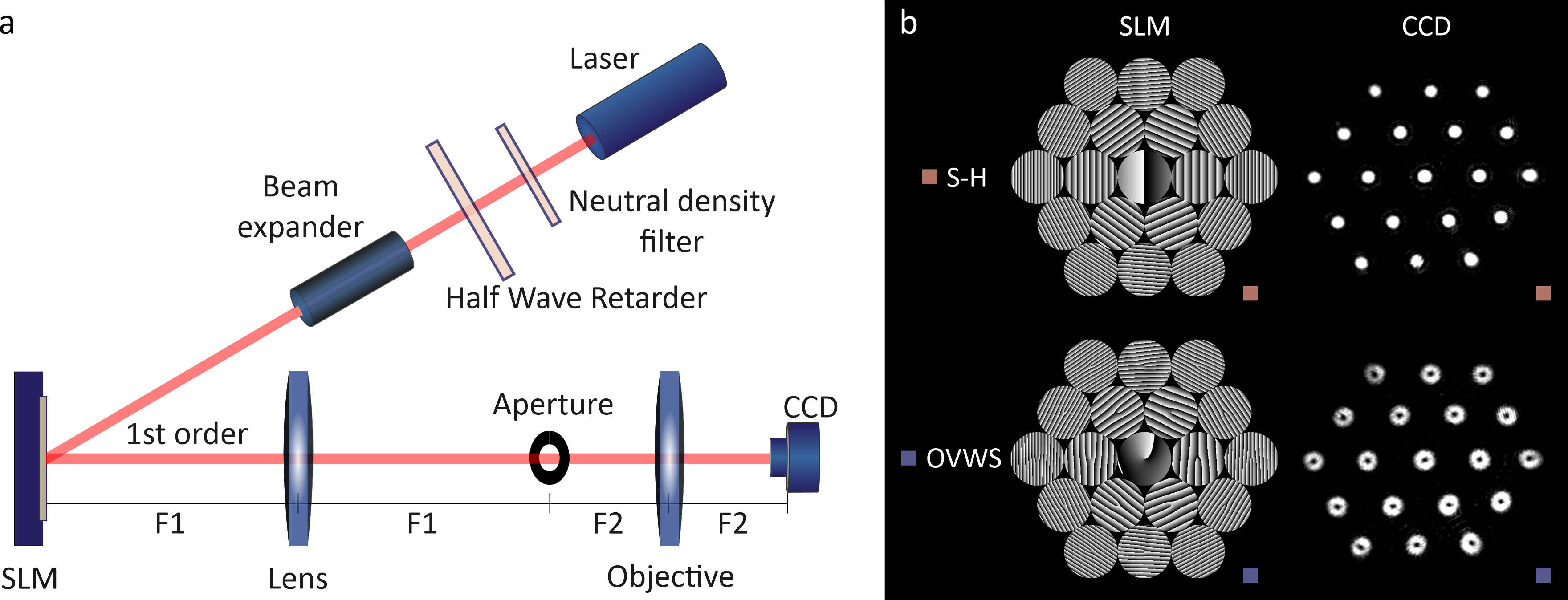}
\caption{a) Experimental setup configuration. b) Exemplary holograms for conventional S-H and OVWS with corresponding intensity distributions.} 
\label{fig:Experimental_setup}
\end{figure}

Figure \ref{fig:Experimental_setup}a shows the experimental setup configuration. An expanded and linearly polarized laser beam ($\lambda=632$ nm) illuminates the SLM (Holoeye LC-R2500, 1280×768 px resolution, 19 $\mu$m pixel size). A neutral density filter adjusts the beam power, while a half-wave plate controls the polarization angle to achieve optimal SLM modulation efficiency.

The illuminating beam is modulated by the sensor's hologram and is further focused at the Fourier plane through a lens with 200 mm focal length. A circular aperture removes unwanted diffraction orders, and the Fourier plane is imaged onto a CCD camera (Basler ace 2 R a2A3840-45ucBAS) using a 10× microscope objective (0.25 NA).

Using this experimental configuration, we evaluated OVWS performance under varying noise conditions by analyzing wavefront retrieval accuracy for different astigmatism levels. The aberration has been manually introduced by superposing the aberration phase map with the sensor's hologram. Assuring proper correction and calibration of the SLM \cite{DeMars:20, Szatkowski2019}.

Figure \ref{fig:Experimental_setup}b shows exemplary holograms for both conventional S-H and OVWS together with their corresponding intensity distributions recorded by the detector. Both configurations are evaluated simultaneously using identical parameters: the number of subapertures (19), subaperture diameter, nx=200, ny=20, and a scaling factor of $s=0.25$. The transition between conventional S-H and OVWS is achieved by adjusting the topological charge value $l$ in equation \ref{eq:phifinal}.

In the experiment, we fixed the camera exposure time to 400 ms and varied the camera gain parameter from 10 dB to 40 dB with 2dB step, registering 100 experimental images for every even gain value. By increasing gain, we amplified all pixel values of the acquired image, effectively increasing the shot noise level. It is worth mentioning that the chosen gain range spans from analog to digital amplification, with 30 dB as the transition value. 

To retrieve the wavefront, the reference was set as the average of 100 images for the medium analog gain in the considered range (gain = 16 dB) without any aberration introduced. This value provided the optimal intensity dynamic range without saturation, while averaging over 100 instances reduced the gain noise impact. In the next step, the retrieved wavefront $\phi_k$ has been calculated by tracking the shift of each spot from its reference position following equations \ref{eq:retrieval} - \ref{eq:residual_variance}.

\begin{figure}[htbp]
\centering
\includegraphics[width=\textwidth]{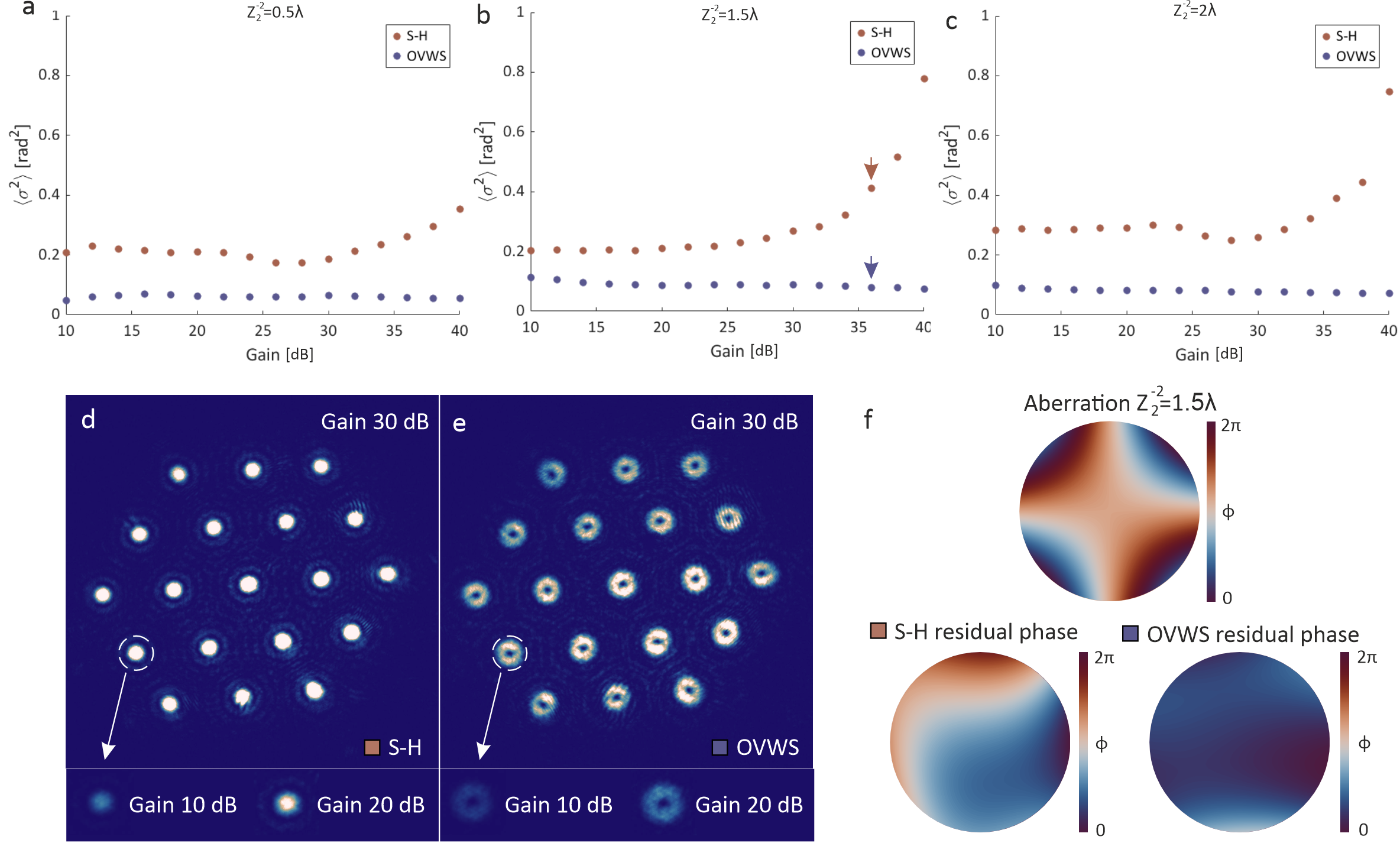}
\caption{Comparison of sensor performance under varying noise conditions. a-c) Mean residual phase variance $\langle\sigma^2\rangle$ as a function of camera gain for three astigmatism levels: a) $0.5\lambda$, b) $1.5\lambda$, and c) $2.0\lambda$. d-e) Intensity distributions from d) conventional S-H sensor and e) OVWS sensor with $Z^{-2}_2=0.5\lambda$ astigmatism at camera gain of 30~dB. 20~dB and 10~dB examples are shown as insets. f) Introduced aberration $Z^{-2}_2=1.5\lambda$ and residual phases from a single ensemble for both conventional S-H and OVWS at 36~dB gain. The arrows in b) mark the corresponding mean values from which this single example was drawn.}
\label{fig:Experimental_results}
\end{figure}

This scheme has been applied to different $\phi_{\text{introduced}}$ patterns containing oblique astigmatism: $Z^{-2}_2=0.5\lambda$, $Z^{-2}_2=1.5\lambda$, and $Z^{-2}_2=2\lambda$ defined for the $d=14.6$ mm pupil. The combined results are shown in Figure \ref{fig:Experimental_results}. 

Figure~\ref{fig:Experimental_results}a-c shows the mean residual phase variance $\langle\sigma^2\rangle$ as a function of camera gain for three astigmatism levels: a) $0.5\lambda$, b) $1.5\lambda$, and c) $2.0\lambda$. In all cases, the OVWS provided lower $\langle\sigma^2\rangle$ values than conventional S-H. Moreover, it remains stable despite increased gain. This is primarily due to the efficiency of the Laguerre-Gaussian transform localization method, consistent with the results reported in~\cite{Szatkowski2022}.

The exemplary intensity distributions for selected gain values are shown in Figure~\ref{fig:Experimental_results}d for the conventional S-H sensor and in Figure~\ref{fig:Experimental_results}e for the OVWS. OVWS offers a higher dynamic range and does not saturate as quickly as the conventional S-H sensor---the peak intensity values are reduced due to optical power distributed over a greater area in the case of OVWS. Figure~\ref{fig:Experimental_results}f shows the introduced aberration $Z^{-2}_2=1.5\lambda$ and residual phases from a single ensemble for both conventional S-H and OVWS visualising the advantage of the OVWS in the noisy regime. 

Both numerical and experimental results demonstrate the practical advantages of the OVWS approach. The OVWS consistently outperformed the conventional S-H sensor across all tested conditions.

\section{Conclusion}

This work introduced the first concept of an Optical Vortex Wavefront Sensor, which augmented the conventional Shack–Hartmann architecture with phase singularities, resulting in improved shot-noise resilience and enhanced accuracy in wavefront phase retrieval. Such modification paved the way for a novel tracking algorithm, enabling high-precision spot detection. To demonstrate this proof of concept, we proposed an SLM-based version of the sensor. 

We evaluated its performance in comparison to its S-H counterpart across a broad SNR range, from low-light to high saturation conditions, each time improving the accuracy of the retrieved wavefront. We evaluated its performance in comparison with the Shack–Hartmann sensor over a wide SNR range, from low-light to high-saturation conditions, with the OVWS demonstrating improved accuracy in wavefront retrieval at all levels.

\begin{figure}[htbp]
\centering
\includegraphics[width=280 pt]{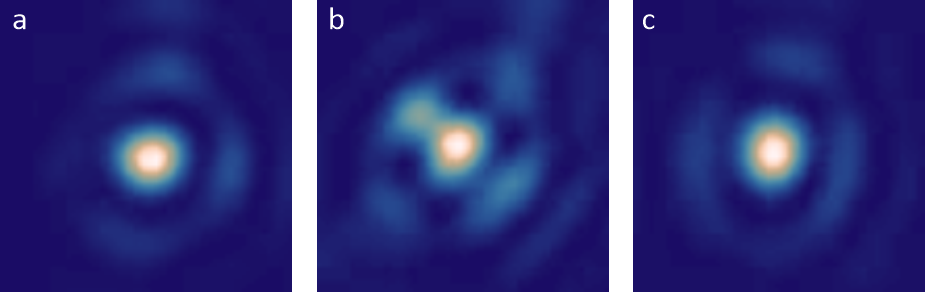}
\caption{a) Experimental image of the Gaussian beam without any aberration in a optical system b) Image aberrated with $Z^{-2}_2=5\lambda$ c) Image after wavefront correction using the OVWS. All images have been registered with the setup presented in Figure \ref{fig:Experimental_setup}a}
\label{fig:Gaussian_image}
\end{figure}

The SLM-based sensor can be efficiently integrated into modern optical setups equipped with spatial light modulators or Digital Micromirror Devices \cite{Cox:21}, enabling correction of basic optical element misalignments. Figure \ref{fig:Gaussian_image} illustrates the experimental results of such a correction, further demonstrating the sensor's versatility in real optical systems.

The most obvious natural approach to development is to proceed with a refractive version of the sensor, which would superimpose a microlens with the spiral phase plate within each subaperture. The performance of the refractive OVWS would strongly rely on the quality of the generated optical vortex, which would be affected by the diameter of the subaperture and the ability to provide a smooth 2$\pi$ phase gradient. Typical subapertures in off-the-shelf solutions are in the range of approximately 150 $\mu$m, which should be large enough to generate an optical vortex of relatively good quality \cite{Nguyen2022}. Both the holographic and the refractive implementations of this concept have been submitted for patent protection \cite{szatkowski2025wavefront}.

\section*{Acknowledgments}
This research was funded by Narodowe Centrum Badań i Rozwoju (0230/L-13/2022) and NSF Directorate for Engineering (2047592).

\section*{Author Contributions}
\textbf{M. Łukowicz}: Investigation (lead); Methodology (equal); Visualization (lead); Validation (supporting); Writing – review \& editing (supporting).
\textbf{A.K. Korzeniewska}: Investigation (supporting); Methodology (equal); Software (supporting); Validation (lead); Writing – review \& editing (supporting).
\textbf{K. Kalinowski}: Software (lead).
\textbf{R. Cichowski}: Methodology (supporting); Visualization (supporting).
\textbf{R. Porras-Aguilar}: Supervision (supporting); Conceptualization (equal); Validation (supporting); Writing – review \& editing (supporting).
\textbf{M. Szatkowski}: Supervision (lead); Conceptualization (equal); Validation (supporting); Writing – original draft (lead).

All authors contributed to the discussion of the results and reviewed the final version of the manuscript.

\section*{Disclosures}
The authors declare no conflicts of interest.

\section*{Data Availability}
Data underlying the results presented in this paper are not publicly available at this time but may be obtained from the authors upon reasonable request.

\bibliographystyle{unsrt}
\bibliography{references}

\section*{Supplemental document}
\label{Supplemental_document}

\section*{S1 Localisation algorithms}
\label{sec:supplement_S1}

We used two localisation algorithms, depending on the type of sensor applied. For the conventional Shack-Hartmann version, the standard weighted centroid localisation algorithm has been utilized:

\begin{itemize}
    \item Record the intensity distribution $I(x,y)$ with the camera sensor.
    \item Identify the region of interest (ROI) containing the spot.
    \item Extract the intensity values within the ROI at pixel coordinates $(x_i, y_i)$ where $i = 1, 2, \ldots, N$ and $N$ is the total number of pixels in the ROI.
    \item Calculate the weighted centroid coordinates using intensity as weights:
    \begin{align*}
    x_c &= \frac{\sum_{i=1}^{N} I(x_i, y_i) \cdot x_i}{\sum_{i=1}^{N} I(x_i, y_i)}, \\
    y_c &= \frac{\sum_{i=1}^{N} I(x_i, y_i) \cdot y_i}{\sum_{i=1}^{N} I(x_i, y_i)}.
    \end{align*}
    \item The spot center is located at $(x_c, y_c)$.
    \item Follow this procedure for subsequent spots in the array.
\end{itemize}

The Optical Vortex Wavefront Sensor produced singular points in the focal spots; therefore, the localisation algorithm has been tailored to take advantage of the singular point existence. Following works \cite{Wang2006, Szatkowski2022}, the procedure can be reduced to the following steps:

\begin{itemize}
    \item Record the intensity distribution $I(x,y)$.
\item Set the bandwidth parameter $\sigma$ based on the expected vortex core size - diameter, where the vortex intensity is half-width.

\item Define the charge-1 Laguerre-Gaussian detection kernel $\mathbf{K}_{\text{LG}}(x,y)$:
    \begin{equation*}
    \mathbf{K}_{\text{LG}}(x,y) = (i\pi^2\sigma^4)(x+iy)\exp(-\pi^2\sigma^2(x^2+y^2)),
    \end{equation*}
    where $i$ is the imaginary unit and $(x,y)$ are pixel coordinates.

\item Convolve the recorded intensity with the detection kernel:
    \begin{equation*}
    \Psi(x,y) = I(x,y) \ast \mathbf{K}_{\text{LG}}(x,y).
    \end{equation*}

\item Extract zero-crossing contours from the real and imaginary parts of $\Psi(x,y)$:
    \begin{align*}
    C_{\text{Re}} &: \text{Re}[\Psi(x,y)] = 0, \\
    C_{\text{Im}} &: \text{Im}[\Psi(x,y)] = 0.
    \end{align*}

\item Locate intersection points of these contours at the $0$ level:
    \begin{equation*}
    (x_v, y_v) : \text{Re}[\Psi(x_v,y_v)] = \text{Im}[\Psi(x_v,y_v)] = 0.
    \end{equation*}

\item Identify the vortex center as the intersection with the lowest recorded intensity $I(x_v,y_v)$.

\item Follow this procedure for subsequent vortex spots in the array.

\end{itemize}

\end{document}